\documentclass[final,leqno]{siamltexmm}
\usepackage{amsmath,amssymb}

\usepackage{graphicx}
\usepackage{dcolumn}
\usepackage{bm}
\usepackage{subeqnar}
\usepackage{overpic,color,rotating}
\usepackage{overpic}
\usepackage{gensymb}

\newcommand{\bx}{\mathbf{x}}
\newcommand{\bF}{\mathbf{F}}
\newcommand{\bA}{\mathbf{A}}
\newcommand{\bX}{\mathbf{X}}
\newcommand{\by}{\mathbf{y}}
\newcommand{\bg}{\mathbf{g}}
\newcommand{\bphi}{\boldsymbol{\phi}}

\graphicspath{{figures/}}

\title{Dynamic Mode Decomposition for Plasma Diagnostics and Validation}

\author{Roy Taylor\thanks{Department of Physics, University of Washington, Seattle, WA.   (email: {rktaylor@uw.edu})  Questions, comments, or corrections to this document may be directed to that email address.} 
\and J. Nathan Kutz\thanks{Department of Applied Mathematics, University of Washington, Seattle, WA. 98195-2420.  }
\and Kyle Morgan\thanks{Department of Physics, University of Washington, Seattle, WA}
\and Brian Nelson\thanks{Department of Electrical Engineering, University of Washington, Seattle, WA}}

\begin{document}
\maketitle
\newcommand{\slugmaster}{%
\slugger{sifin}{}{}{}{}}

\begin{abstract}
We demonstrate the application of the Dynamic Mode Decomposition (DMD) for the diagnostic analysis of the nonlinear dynamics of a magnetized plasma in resistive magnetohydrodynamics.  The DMD method is an ideal spatio-temporal matrix decomposition that correlates spatial features of computational or experimental data while simultaneously associating the spatial activity with periodic temporal behavior.  DMD can produce low-rank, reduced order surrogate models that can be used to reconstruct the state of the system and produce high-fidelity future state predictions. This allows for a reduction in the computational cost, and, at the same time, accurate approximations of the problem, even if the data are sparsely sampled.  We demonstrate the use of the method on both numerical and experimental data, showing that it is a successful mathematical architecture for characterizing the HIT-IS magnetohydrodynamics.  Importantly, the DMD decomposition produces interpretable, dominant mode structures, including the spheromak mode and a pair of injector-driven modes.  In combination, the 3-mode DMD model produces excellent dynamic reconstructions that can be used for 
future state prediction and/or control strategies.
\end{abstract}



\section{Introduction}

Carbon-free, high-availability electric power generation would greatly benefit from an economical fusion reactor. The self-organized spheromak magnetic confinement configuration\cite{spher_rev} offers a path towards such a goal. Magnetic helicity, the linkage of magnetic flux with flux, is the best constant  of motion for a magnetized plasma, decaying only on the resistive diffusion time. Relaxation of the magnetic field energy is much more rapid than the resistive diffusion time, thus magnetic energy decays toward a minimum energy state,  conserving helicity. The end product, known as a Taylor state \cite{spher_rev} is compact, self-organized, and simply-connected, which assumes a configuration determined by its helicity-conserving (conductive) boundary. Helicity can be ``injected'', overcoming resistive decay, by locally driving current along magnetic flux, and subsequent relaxation processes distribute current throughout the spheromak volume.\cite{spher_rev} The helicity injected torus with steady inductive helicity injection (HIT-SI)\cite{hitsi} experiment studies formation and steady-state sustainment of a spheromak, and it is with this experiment that we demonstrate the capacity of the Dynamic Mode Decomposition for analysis and diagnostics. 

The HIT-SI experiment produces helicity injection current drive in a steady-state inductive manner.\cite{hit_sust} Furthermore, recent results have found that the current drive relaxation processes can be {\em imposed}\/ by non-axisymmetric perturbations, rather than relying on plasma instabilities redistributing the driven current. This allows formation and sustainment of a {\em stable}\/ spheromak object using a process called imposed dynamo current drive (IDCD).\cite{idcd} IDCD is a very efficient method of current drive, which has been used to design the ``dynomak'' reactor concept.\cite{dynomak} The compact nature of the spheromak results in a 1000~MWe dynomak reactor with an overnight capital cost less than a comparable power output coal-fired power plant.

Numerical simulations of the magnetohydrodynamics in the spheromak geometry provide a critically enabling tool for engineering design and parameter studies, especially as important plasma diagnostics are computed.
Modern numerical simulations of magnetohydrodynamic systems necessarily output significantly larger data sets than those generated by the limited sensors monitoring the experimental dynamics.   Specifically, diagnostic apparatus for plasma experiments are limited to a number of spatial sensor locations along the spheromak.  As a result, two central problems emerge: (a) how to parse physically meaningful structure from limited (sparse) number of diagnostic measurements, and (b) how to leverage diagnostics to validate numerical simulations and improve the general reliability of modeling (therefore overcoming deficiencies in experimental measurement). 

In addressing these two problems, it is customary to employ simple data mining techniques to reveal low-rank, physically interpretable spatiotemporal patterns in the spheromak.  In 1993 de Wit et al. introduced the biorthogonal decomposition as a highly extensible, multivariate approach to interpreting plasma data sets \cite{bd}. While the biorthogonal decomposition, BOD for short, offers powerful insight into the structure of spatially-extended time series measurements, it is still limited in its ability to resolve dynamical properties or use low-rank information for modeling. In this paper we consider an extension of BOD analysis, the \textbf{dynamic mode decomposition} (DMD), for the purpose of diagnostics and validation.

Whereas biorthogonal decompositions resolve modes in space and time separately, the DMD modes preserve coherent spatiotemporal patterns, and subsequently DMD spectral quantities parametrize the time and space dynamics themselves and are ranked by energy, as with BOD modes. Furthermore the DMD relaxes the constraint that modes must be mutually orthogonal, which proves useful for systems where, with dimensionality reduction, physically meaningful structures are not necessarily orthogonal. To demonstrate the utility of these differences, we apply the dynamic mode decomposition to a magnetically confined plasma. In particular we consider the application of DMD analysis to the HIT-SI spheromak's surface magnetic probe measurements. For diagnostic applications we consider the dynamic modes emergent from just the surface probe data.  For validation, we compare these modes to the mode structure emergent from NIMROD \cite{sovinec04} magnetohydrodynamics simulations. Finally we consider the extensibility of DMD modeling to future-state prediction with implications for cross-validation and, potentially, controls.  Importantly, the DMD method provides an interpretable, low-rank set of modes that can be directly linked with experimental diagnostic recordings, thus enhancing the connection between theory and experiment.

The paper is outlined as follows:  In Sec.~2, the geometry and experimental configuration of
the spheromak device is given.  This highlights the HIT-SI architecture used for the modeling efforts.
In Sec.~3, a review is given of the model reduction techniques that can be used for 
diagnostic analysis, including the commonly used BOD decomposition as well as the DMD reduction
introduced in this work.  Section~4 highlights the application of the DMD method to the simulation 
data, demonstrating the efficacy of the method and the highly interpretable, low-rank  spatiotemporal
patterns computed.  The resulting decomposition is also validated and tested for its ability to accurately produce
future-state predictions.  Section~5 closes the manuscript by summarizing the achievements of the DMD
decomposition and outlining various future studies possible with the DMD architecture.

\begin{figure}[t]
\centering
\includegraphics[scale=.24]{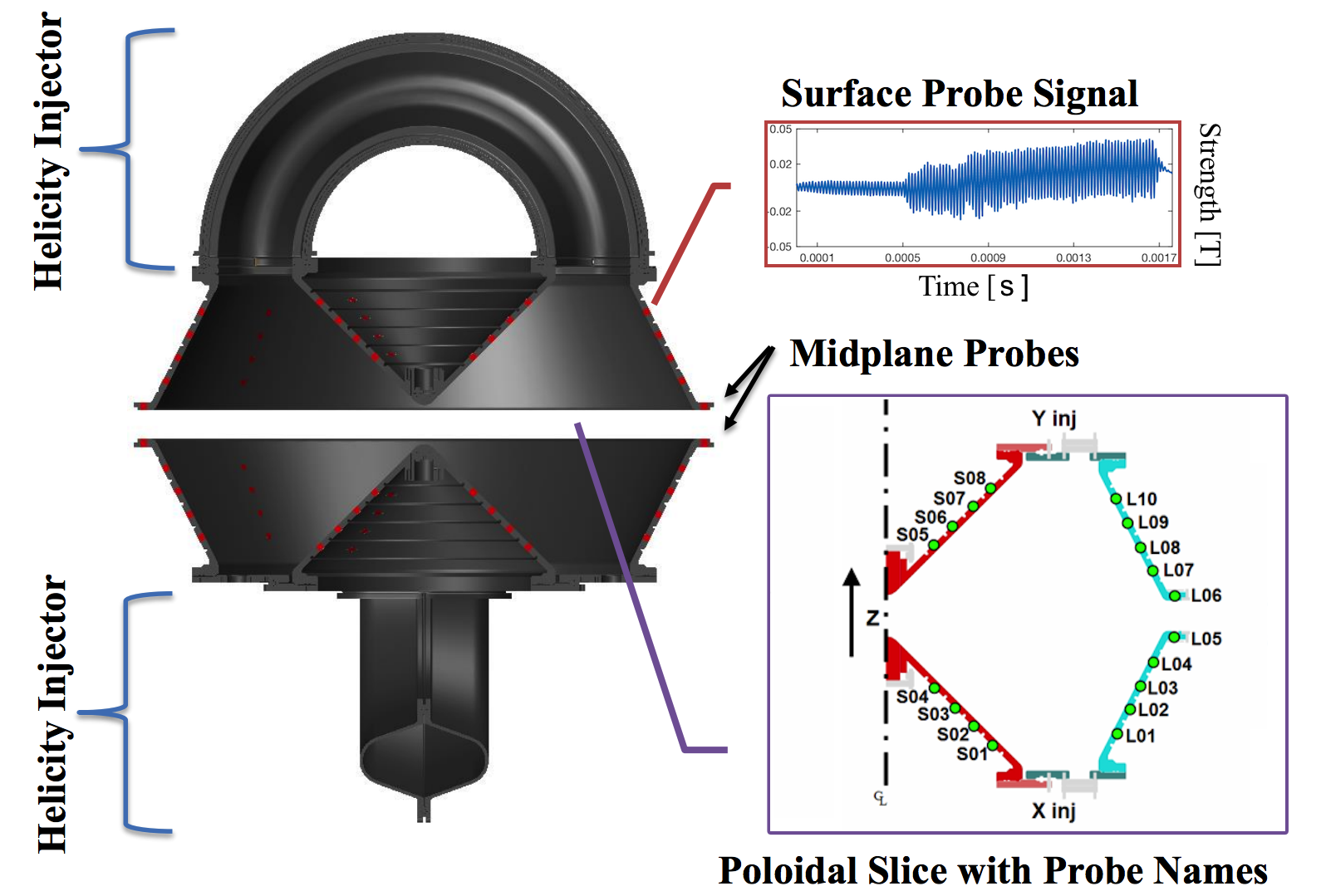}
\caption{Magnetic probes (red dots) wrap around four poloidal cross-sections of the confinement volume, categorized as \emph{midplane} (probe locations L06 and L07) and \emph{surface} on the inner \emph{small} cone (e.g.\ S08) or the outer \emph{large} cone (e.g.\ L10) as enumerated on the bottom-right cut-away. Each probe produces its own independent time series signal (top right). The helicity injectors sit at either end of the confining volume.}
\label{fig:probes}
\end{figure}

The HIT-SI device is a current drive experiment for magnetic confinement fusion \cite{hitsi}.  The device consists of a simply-connected bow-tie shaped confinement region and two inductive helicity injectors, as seen in 
Fig.~\ref{fig:probes}.  Each injector has two sets of coils to inductively drive parallel magnetic flux and electric current and to inject magnetic helicity $(K = \int A \cdot B \mathrm{d}^3x)$ into the confinement region.  Due to the inductive nature of the drive, the injectors must be operated in an oscillatory manor to allow steady state operation.  This is parametrized by the loop voltage $(V_{inj} = V_0 sin(\omega t))$ and an injector flux $(\psi_{inj} = \psi_0 sin(\omega t))$.  The helicity injected is given by $\dot{K}_{inj} = 2 V_{inj} \psi_{inj}$, so that the two injectors are driven $90^\circ$ out of phase in order to cause constant helicity injection. Initially, a state connecting the injectors to the central region is driven, but after a short time, a relaxation event occurs which transitions the plasma towards the minimum energy state, an axisymmetric spheromak.  The helicity decays due to resistivity, but it is replenished by the injectors, leading to a sustained steady state equilibrium.  Additionally, HIT-SI is enclosed by a copper wall that is coated with an insulating layer to create the boundary conditions ${\bf B}_\perp = 0$ and ${\bf J}_\perp = 0$ on the timescale of a discharge. Certain control parameters are chosen in advance of each shot, such as injector phasing and Steady Inductive Helicity Injection \cite{hitsi_operation} frequency (SIHI), which will become important for this analysis.

\subsection{Surface Probes and Diagnostics}

The HIT-SI experiment operates with magnetic field probes encircling four poloidal cross sections at toroidal angles 0, 45, 180, and 225 degrees. These probes come in two sets: sixteen \emph{surface} probes and two additional \emph{midplane} probes along the diagnostic gap, as shown in Fig.~\ref{fig:probes}. Each of these produce three time series signals, for the magnetic field in three component directions. For the sake of this paper, we are concerned with the \textbf{poloidal} component of the magnetic field signals, as shown in the top-right of Fig.~\ref{fig:probes}. Additionally we constrain ourselves to the \textbf{surface probes} located at \textbf{toroidal angle 225 degrees} only.

\begin{figure}[t]
\centering
\includegraphics[scale=.11]{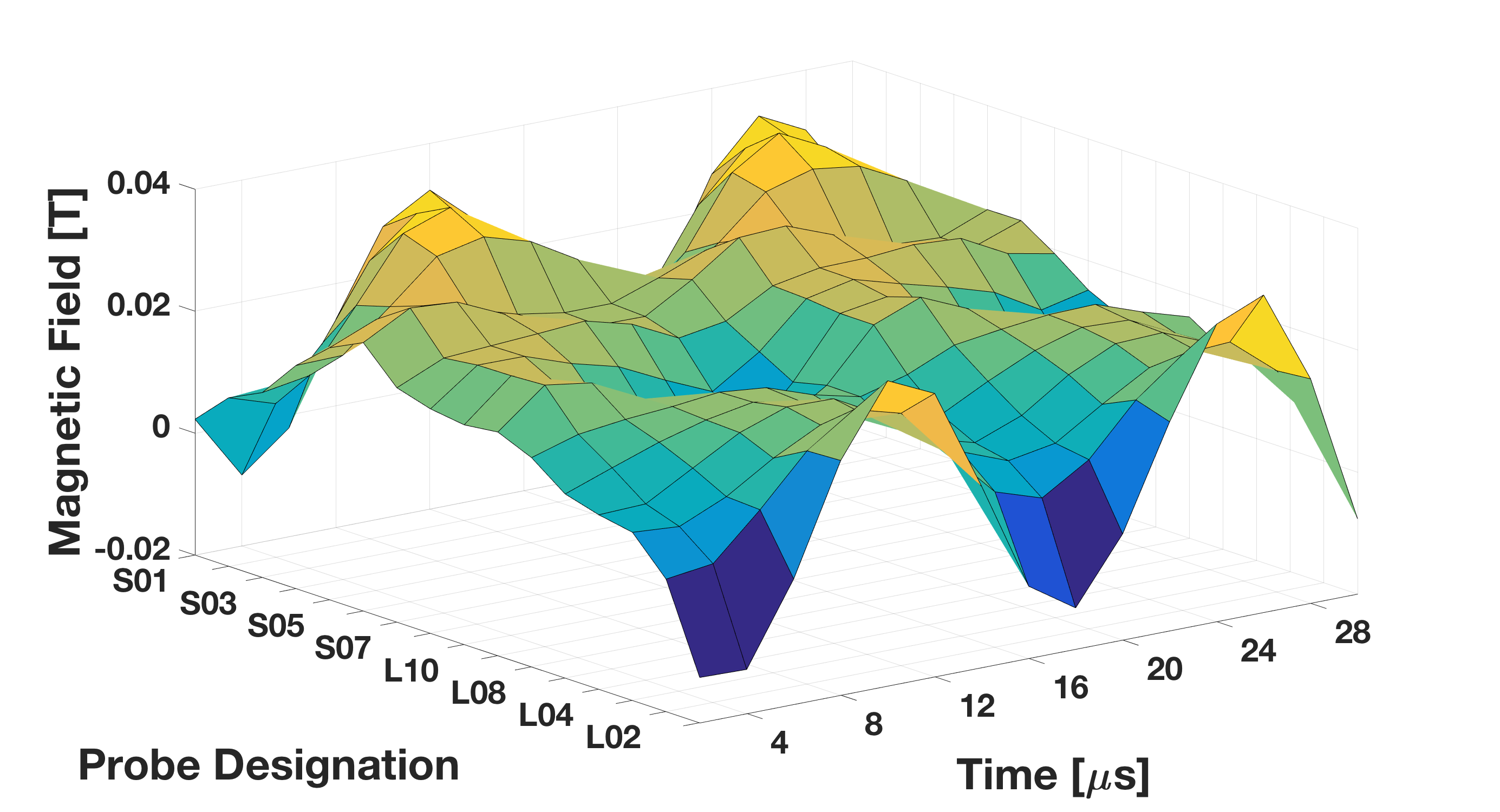} \\
\caption{Time series signals for each probe (as in FIGURE \ref{fig:probes}) are stacked as rows in a state matrix, displayed as a surface for demonstration of temporal and spatial characteristics.}
\label{fig:time series}
\end{figure}

To visualize spatial-temporal patterns in the magnetic field, we arrange each of the probe's time series data into a surface plot, as shown in Fig.~\ref{fig:time series}. This gives an intuitive understanding of modal properties produced by the experiment so that we can compare it with the DMD analysis. For experimental data, probe measurements are resolved at two microsecond time steps. (The NIMROD simulation is slightly higher resolution, at 1.2 microsecond time steps.)

\subsection{NIMROD Modeling}

HIT-SI can be simulated using the \textit{Non-Ideal Magnetohydrodynamics with Rotation - Open Discussion} (NIMROD) code \cite{sovinec04}.  NIMROD solves the equations of MHD in 3-spatial dimensions assuming a fully ionized Deuterium plasma:
\begin{subeqnarray}
\frac{\partial  n}{\partial t} &=& \nabla \cdot { n {\bf V}} + D\nabla^2{\ n} \\
\frac{\partial {\bf V}}{\partial t} + \left( {\bf V} \cdot \nabla \right) {\bf V} &=& \frac{1}{\rho} \left( {\bf J} \times {\bf B}  - \nabla p + \nabla \cdot \Pi \right) \\ 
\frac{ n}{\gamma - 1} \left( \frac{\partial}{\partial t} + {\bf V} \cdot \nabla \right)  T_i &=& -\nabla \cdot {\bf V} - \Pi : \nabla {\bf V} - \nabla \cdot {\bf q} \\
\frac{\partial {\bf B}}{\partial t} &=& - \nabla \times {\bf E} \\
{\bf E} &=& \left( -{\bf V} \times {\bf B} + \eta {\bf J} + \frac{{\bf J} \times {\bf B}}{ne} + \frac{m_e}{ne^2} \frac{\partial {\bf J}}{\partial t} \right)
\end{subeqnarray}
Where ${\bf V}$ is the center-of-mass velocity, ${\bf B}$ is the magnetic inductance, ${\bf E}$ is the electric field, $n$ is the particle density, and $ T_i$ is the ion temperature.  It is assumed that the electron temperature is uniform in space and constant in time, at $T_e = 12$ eV.  Limited calculations have been performed allowing evolution of $T_e$ which support this assumption, though a full simulation has thus far proved computationally intractable.  This assumes that the thermal losses from thermal conduction to the boundaries of the device are equal to the Ohmic heating power of the plasma current.  Additionally, the following constitutive relations  hold
\begin{subeqnarray}
&& \rho = m_i n \\
&& {\bf J} = \frac{1}{\mu_0} \nabla \times {\bf B} \\
&& \Pi = -\rho \nu \left( \nabla {\bf V} + \nabla {\bf V}^T - \frac{2}{3} \nabla \cdot {\bf V} \right) \\
&&  p = { n} k_B{ T_i} \\
&& {\bf q} = \left( k_\parallel \hat{b} \hat{b} + k_\perp \left( I - \hat{b} \hat{b} \right) \right) \cdot \nabla { T_i} 
\end{subeqnarray}

\noindent For toroidal magnetic devices such as HIT-SI, NIMROD solves the equations in cylindrical coordinates.  The spatial discretization of the code is a combination of Fourier decomposition in the azimuthal direction and a finite-element representation in the \textit{R-Z} plane.  Typically NIMROD is successful in modeling HIT-SI with a somewhat coarse ($n_{modes} = 11$) angular resolution and a fine resolution ($m_r = m_z = 24$ with polynomial degree of 4) for the \textit{R-Z} plane.
A particle diffusivity $D = 1000$ m$^{-2}$ is used for numerical stability of both the Hall term in the magnetic advance and the continuity equation.  
The resistivity $\eta = 1.1 \times 10^{-5}$ $\Omega$m is equivalent to values used in previous zero-$\beta$ modeling of HIT-SI \cite{hit_akcay}\cite{hansen2} to obtain agreement with experimental results.  The anisotropic thermal conduction coefficients used in the model are computed from Braginskii \cite{braginskii} from spatially dependant $\bf B$, $T_i$, and $n$.   A viscosity of $\nu = 300$ m$^2$/s, calculated from Braginskii parallel viscosity assuming $T_i=12$ eV and $n=1.5\times 10^{19}$m$^{-3}$ is used.

\begin{figure}[t]
\centering
\includegraphics[width=0.5\textwidth]{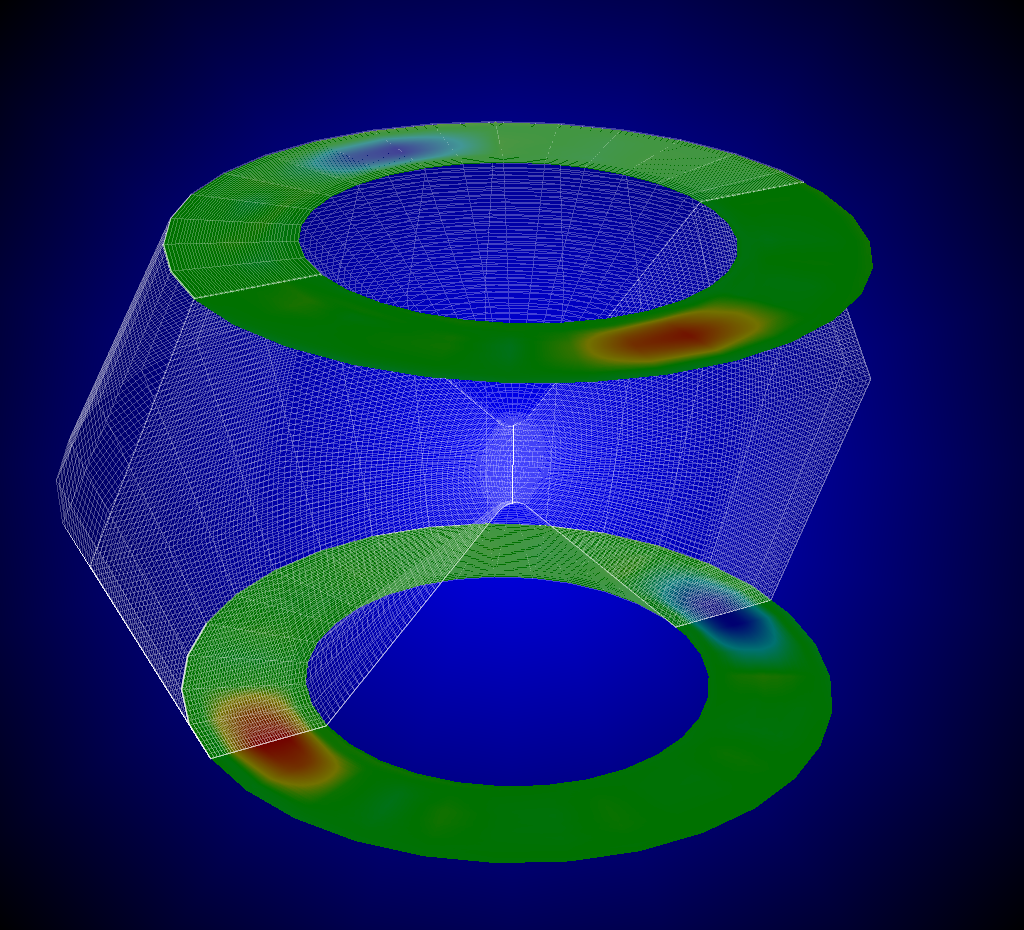}
\caption[The NIMROD mesh]{The mesh used for NIMROD simulations of HIT-SI.  NIMROD requires a toroidally symmetric grid, so the injectors must be represented as boundary conditions.  The pseudocolor represents the ${\bf B}_{\perp}$ used for the flux injection boundary condition.  A ${\bf E}_\parallel$ condition is similarly used to create an analogous ${\bf J}_\perp$.  A thin layer of high resistivity ${\eta_{edge}} / {\eta_{plasma}} \sim 10^5$ is used to simulate the insulating wall of the experimental device.}
\label{fig:nim_mesh}
\end{figure}

The 3-D nature of the HIT-SI injectors requires some approximations when modeled with the 2-D NIMROD grid.  The computational grid of the NIMROD simulation is strictly the central confinement region with the injector fields modeled as boundary conditions.  The Grad-Shafranov equation is solved on the injector geometry and projected as a combination of ${\bf B}_{\perp}$ and ${\bf E}_\parallel$ to generate $\psi_{inj}$ and $I_{inj}$ matching an experimental discharge.  Elsewhere the boundary is assumed to be perfectly conducting, such that ${\bf B}_\perp = 0$ and ${\bf E}_\parallel = 0$. Figure~\ref{fig:nim_mesh} shows our configuration.  A high-$\eta$ edge layer ($\eta_{edge} \sim 10^5 \eta_{plasma}$) is included to enforce the pseudo-boundary condition ${\bf J}_\perp = 0$.  A detailed description of these boundary conditions and previous comparisons with experimental results can be found in \cite{hit_izzo} \cite{hit_akcay} \cite{hit_bd} and \cite{hansen2}.

\section{Reduced-order Modeling for MHD Systems}

Model order reduction is now becoming common across many areas of the engineering, physical and biological
sciences.  The reduced order models generated, often referred to as surrogate models, aim to capitalize on
low-dimensional structures observed in large scale simulations~\cite{benner2013survey}.  A number of
reductions techniques are possible, and two are highlighted here:  {\em Biorthogonal decomposition}, which is commonly used  in spheromak analysis, and the {\em Dynamic Mode Decomposition} which is our specific innovation for model reduction. \cite{hit_bd}

Matrix decompositions are critically enabling algorithms for diagnostic analysis and scientific computing applications across every field of the engineering, social, biological, and physical sciences.  Of particular importance is the singular value decomposition (SVD), which provides a principled method for dimensionality reduction and computation of interpretable subspaces within which the data reside.  So widespread is the usage of the SVD algorithm, and minor modifications thereof, that it has generated a myriad of names across various communities, including Principal Component Analysis (PCA)~\cite{Pearson:1901}, the Karhunen-Lo\`eve (KL) decomposition, Hotelling transform~\cite{hotellingJEdPsy33_1}~\cite{hotellingJEdPsy33_2}, Empirical Orthogonal Functions (EOFs)~\cite{eof1}, Proper Orthogonal Decomposition (POD)~\cite{HLBR_turb}, and in particular to the fusion community, the Biorthogonal Decomposition ~\cite{bd}.  In each of these cases, the low-rank features extracted from the matrix factorization help provide interpretable, spatially correlated structures that can help inform understanding and potential control protocols.


\subsection{Biorthogonal Decomposition in Theory and Practice}

Numerical simulations are capable of characterizing to high-resolution the spatial dynamics of the spheromak physics.  In contrast, the HIT-SI experimental configuration only allows for a limited number of spatial sensor measurements, thus not allowing for highly-resolved spatial recordings.   As such, validation of the physics-based simulation is imperative.  Validation metrics have been developed using Biorthogonal Decomposition (BOD) to check the accuracy of these simulations \cite{hit_bd,hansen2}. With these metrics, the simulation is considered validated when it consistently agrees with experiment within the repeatable range of the experiment.

The biorthogonal decomposition looks at the eigenfunctions produced by a singular value decomposition (SVD) of experimental magnetic probe signals and compares the structure of the dominant modes.  A typical SVD of HIT-SI's surface probe array produces 3 modes that comprise the majority of the signal energy.  The first mode is the spheromak, while the second and third are a set of orthonormal oscillating modes oscillating at $f_{inj}$, representing the injector modes.  The fourth and fifth modes are second harmonics of the injectors, but are very small, and higher modes all appear to be consistent with signal noise.  The BOD validation metrics compare the shape of these orthonormal eigenfunctions of the signal while weighting their importance by the energy in the given mode.  

BOD validation in its current construction is limited, however. First, biorthogonal techniques, as the name suggests, force an orthonormality onto data that may not be necessary or physically valid. Second, the rigid separation of modes into temporal and spatial independent structures may destroy coherent temporal-spatial patterns of interest for analysis. More generally, application of BOD requires delicate precision:  (i) How can we be certain where to truncate `signal noise'? (ii) How can we be certain of the dynamics for each mode, if the decomposition window contains a transition in the physical system (e.g.\ relaxation)?  These questions motivate the particular application of the \textbf{dynamic mode decomposition} with \textbf{Gavish-Donoho optimal hard thresholding} ~\cite{gavish} advocated herein.

\subsection{Dynamic Mode Decomposition}

The dynamic mode decomposition (DMD) is a matrix factorization method based upon the SVD algorithm.  However, in addition to performing a low-rank SVD approximation, it further performs an eigendecomposition on the computed subspaces in order to extract critical temporal features.  Thus the DMD method provides a spatio-temporal decomposition of data into a set of dynamic modes that are derived from snapshots or measurements of a given system in time (arranged as column state-vectors).   The mathematics underlying the extraction of dynamic information from time-resolved snapshots  is closely related to the idea of the Arnoldi algorithm~\cite{Schmid2010jfm}, one of the workhorses of fast computational solvers.   The data collection process involves two parameters:
\begin{subeqnarray}
  &&  n = \mbox{number of spatial points saved per time snapshot} \nonumber \\
  &&  m= \mbox{number of snapshots taken} \nonumber
\end{subeqnarray}
The DMD algorithm was originally designed to collect data at regularly spaced intervals of time, as in \eqref{eq:DataCollection}.  However, new innovations allow for both sparse spatial~\cite{Brunton2015jcd} and temporal~\cite{Tu:2014b} collection of data as well as irregularly spaced collection times.   Indeed, Tu \emph{et al.}~\cite{Tu2014jcd} provides the most modern definition of the DMD method and algorithm.\\

\noindent {\bf Definition:  Dynamic Mode Decomposition} (Tu \emph{et al.} 2014~\cite{Tu2014jcd}): {\em Suppose we have
a dynamical system (\ref{eq:U}) and two sets of data
\begin{subeqnarray}
 && {\bf X} = \begin{bmatrix}
\vline & \vline & & \vline \\
\bx_1 & \bx_2 & \cdots & \bx_{m-1}\\
\vline & \vline & & \vline
\end{bmatrix} \\
&&\nonumber\\
 && {\bf X}' = \begin{bmatrix}
\vline & \vline & & \vline \\
\bx'_1 & \bx'_2 & \cdots & \bx'_{m-1}\\
\vline & \vline & & \vline
\end{bmatrix}
\end{subeqnarray}
so that $\bx'_k=\bF(\bx_k)$ where $\bF$ is the map in \eqref{eq:FlowMap} corresponding to the evolution of \eqref{eq:U} for time $\Delta t$.   DMD computes the leading eigendecomposition of the best-fit linear operator $\bA$
relating the data $\bX'\approx\bA\bX$\,:
\begin{equation}
  {\bf A} = {\bf X}' {\bf X}^\dag.
  \label{eq:DMD}
\end{equation}
The DMD modes, also called dynamic modes, are the eigenvectors of $\bA$, and each DMD mode corresponds to a particular eigenvalue of $\bA$.}

In the DMD architecture, we typically consider data collected from a dynamical system
\begin{equation}
  \frac{d{\bf x}}{dt} = {\bf f}({\bf x},t;{\boldsymbol{\mu}}) \, ,
  \label{eq:U}
\end{equation}
where ${\bf x}(t)\in\mathbb{R}^n$ is a vector representing the state of our dynamical system at time $t$, $\boldsymbol{\mu}$ contains parameters of the system, and ${\bf f}(\cdot)$ represents the dynamics.  For instance, the state vector ${\bf x}$ denotes the
surface magnetic probe measurements after numerical discretization in our specific example while $\boldsymbol{\mu}$ contains inputs/control knobs for the system, such as the SIHI frequency or injector phasing.
The state $\bx$ is typically quite large, having dimension $n\gg 1$.  This is typically required for producing high-fidelity and well-resolved simulations of the plasma dynamics.
Measurements of the system
\begin{eqnarray}
\by_k = \bg(\bx_k),
\end{eqnarray}
are collected at times $t_k$ from $k=1,2,\cdots,m$ for a total of $m$ measurement times.
The measurements are typically the same plasma state parameters as before, so that $\by_k=\bx_k$, however, the DMD architecture
allows for a more nuanced viewpoint of observables.  This is beyond the scope of the current work, but such ideas
are related to Koopman theory~\cite{DMDbook} and may be extensible to controls in the future.

The DMD framework takes an equation-free perspective where the original, nonlinear dynamics (e.g.\ plasma field kinetics) may be unknown.  Thus data measurements of the system alone are used to approximate the dynamics and predict the future state.    The DMD procedure  constructs the proxy, approximate locally linear dynamical system
\begin{equation}
  \frac{d {\bf x}}{dt} = {\large \bA c}{\bf x}\label{eq:SSLinCont}
\end{equation}
with initial condition ${\bf x}(0)$ whose well-known solution is
\begin{equation}
   {\bf x}(t)=  \sum_{k=1}^n  \bphi_k \exp(\omega_k t) b_k=\bphi \exp(\boldsymbol{\Omega} t)\mathbf{b}\,
   \label{eq:omegaj}
\end{equation}
where $\bphi_k$ and $\omega_k$ are the eigenvectors  and eigenvalues of the matrix ${\bf A}$, and the coefficients $b_k$ are the coordinates of $\bx(0)$ in the eigenvector basis.

The DMD algorithm produces a low-rank eigen-decomposition \eqref{eq:EigenDisc} of the matrix $\bA$ that optimally fits the measured trajectory $\bx_k$ for $k=1,2,\cdots,m$ in a least square sense so that
\begin{equation}
\|\bx_{k+1}-\bA\bx_k\|_2 \label{eq:snapshoterror}
\end{equation}
is minimized across all points for $k=1,2,\cdots, m-1$.
The optimality of the approximation holds only over the sampling window where ${\bf A}$ is constructed, and the approximate solution can be used to not only make future state predictions, but also to derive dynamic modes critical for diagnostics.  Indeed, in much of the literature where DMD is applied, it is primarily used as a diagnostic tool.  This is much like POD analysis where the POD modes are also primarily used for diagnostic purposes.  Thus the DMD algorithm can be thought of as a modification of the SVD architecture which attempts to account for dynamic activity of the data.  The eigendecomposition of the low rank space found from SVD enforces a Fourier mode time expansion which allows one to then make spatio-temporal correlations with the sampled data.

\begin{figure}[t]
\centering
\includegraphics[scale=0.18]{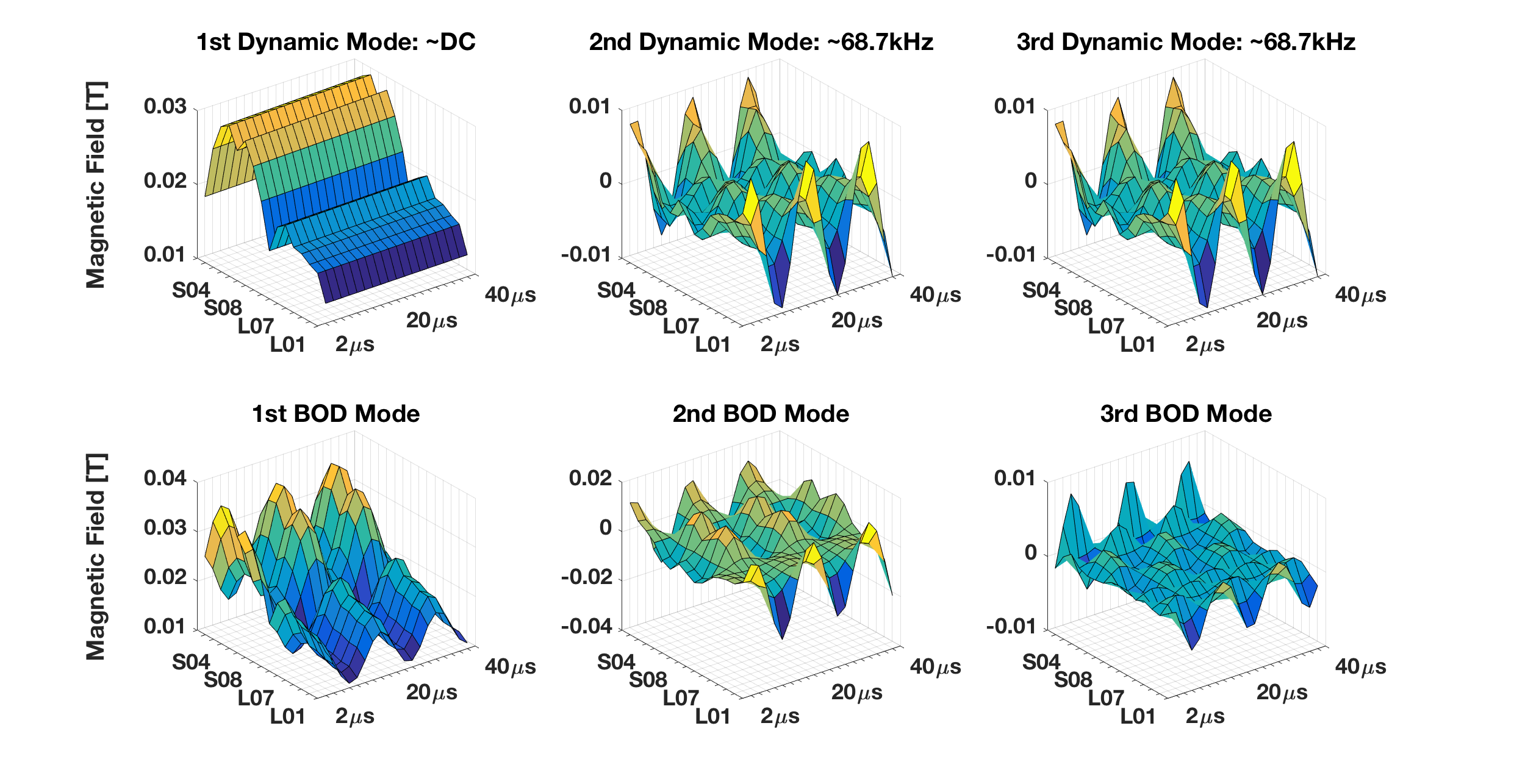}\\
\caption[DMD mode figures]{The DMD does not preserve the orthogonality of BOD modes, but it does resolve coherent temporal-spatial structures with degenerate pairs for oscilliatory modes. The 1st Dynamic Mode is a DC profile matching that of a spheromak in Taylor-state equilibrium, while the 2nd/3rd Dynamic Mode pairs produce oscillations at the SIHI frequency (top). In contrast, the dominant BOD mode largely captures the DC profile, but not separately from oscillations at the injector frequency: a symptom of mode-mixing.}
\label{fig:DMD_modes}
\end{figure}

\section{DMD Diagnostic Results}

A full experimental shot, or its corresponding numerical simulation, will measure signal outputs over changing dynamical regimes. As a result, it is nonsensical to apply the DMD--or any data mining technique, such as POD--to the data set \emph{as a whole}. Furthermore, encountering the constraint that DMD convergence does best with $n \gg 1$ and a temporal dimension that is not too much larger than the spatial dimension, we have limited our analysis to \emph{a sliding window of 20 temporal measurements}, an optimal value which we have found yields the most consistent results. At each step forward of the sliding window, we record two DMD spectral quantities:

\begin{enumerate}
\item \textbf{Magnetic energy percentage for the spheromak mode}, defined as the ratio of $b_{spheromak}^2$ to the total $\left(\sum_{k=1}^n b_k^2 \right)$.   As SVD algorithms produce unitless modes, units are therefore carried by the weights $b_k$ as in Eq.~(\ref{eq:omegaj}). Therefore $b_k$ carries units of Tesla, and as $u_{B} = \frac{B^2}{2\mu_0}$, we define `spheromak percentage' as the ratio of the spheromak singular value squared to the sum of squares of the \textbf{full-rank} singular value spectrum. 
\item \textbf{Mode frequency content}, defined as $f = \Im [\omega] / 2\pi$. As the DMD omega value, from Eq.~(\ref{eq:omegaj}), yields a complex number whose real part gives growth/decay and whose imaginary part gives oscillation, it is trivial to track the frequency content \textbf{ranked by mode energy} as the sliding window progresses. (This is also how we track which mode is the spheromak, i.e., the DC mode.)
\end{enumerate}

Beyond spectral quantities, it is important to note how the data reconstructs the spatial-temporal dynamics given the dominant DMD modes.  In particular, we wish to characterize how this improves upon the biorthogonal decomposition technique.  In Fig.~\ref{fig:DMD_modes}, the difference between dynamic modes and biorthogonal modes is highlighted. The DMD produces clean separation of static temporal-spatial structures (the spheromak) from dynamic temporal-spatial structures (the injector-driven dynamics), a separation lacking in traditional BOD reconstructions.  This is by design as the DMD allows for each mode to be associated with a time dynamics.

\subsection{Rank-3 Model and Model Error}

\begin{figure}[t]
\centering
\includegraphics[scale=0.17]{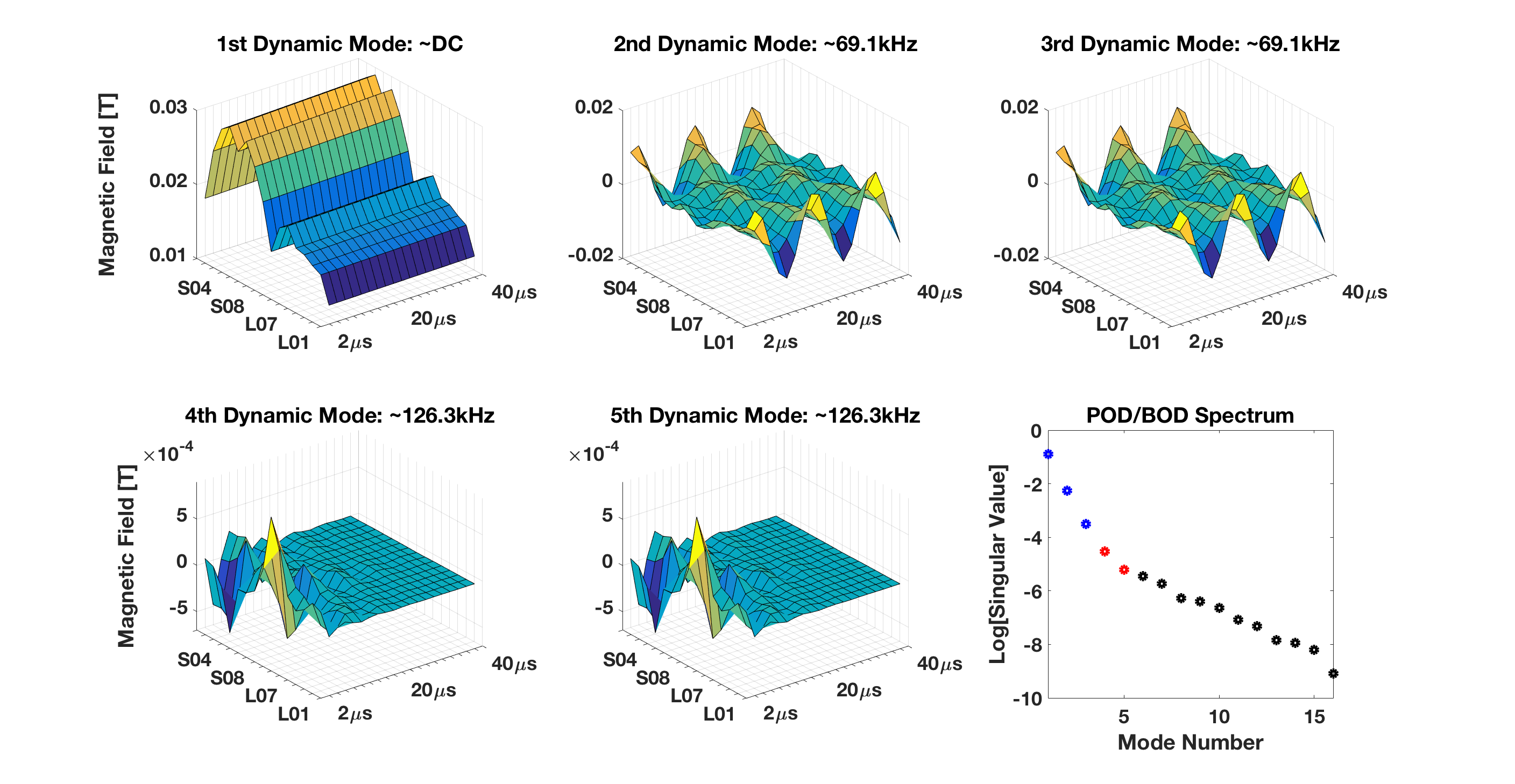} \\
\caption{Examined shots demonstrate true Rank-3 sparsity. Inclusion of the 4th and 5th modes produces rapidly-decaying transients with no physical significance. We therefore conclude that, for a stable HIT-SI plasma, poloidal B measurements are only Rank 3, and that any higher-rank representations produce spurious modes from noise. NB: axis labels are identical to those in FIGURE \ref{fig:DMD_modes}, omitted for visual brevity.}
\label{fig:DMD_comparison}
\end{figure}

Methodologically, we consider both data sets--experimental and NIMROD--under the `soft' constraint of Gavish-Donoho optimal hard thresholding ~\cite{gavish}. That is, the thresholding limits given by Gavish and Donoho assume the data are comprised of a low-rank signal with additive white noise.  The noise statistics in experiment and in our own simulations remain uncharacterized.  However, the Gavish-Donoho hard threshold provides a principled heuristic truncation which is superior than a simple energy threshold.  Applying the Gavish-Donoho metric to a selection of window locations and sizes, it becomes clear that three modes always survive, whereas the fourth and fifth modes only sometimes survive. Therefore we use a three mode decomposition and show this is a good choice when comparing a rank-3 model versus a rank-5 model for both NIMROD and experimental data sets.  As shown in Fig.~\ref{fig:DMD_comparison}, allowing the reconstruction to generate five modes in effect produces spurious modes which rapidly decay away. In fact, these transient modes are apparent at nearly every point along the sliding window, ergo they may be regarded as artifacts of noise inclusion and not, in effect, physical structures.

\begin{figure}[t]
\centering
\includegraphics[scale=.2]{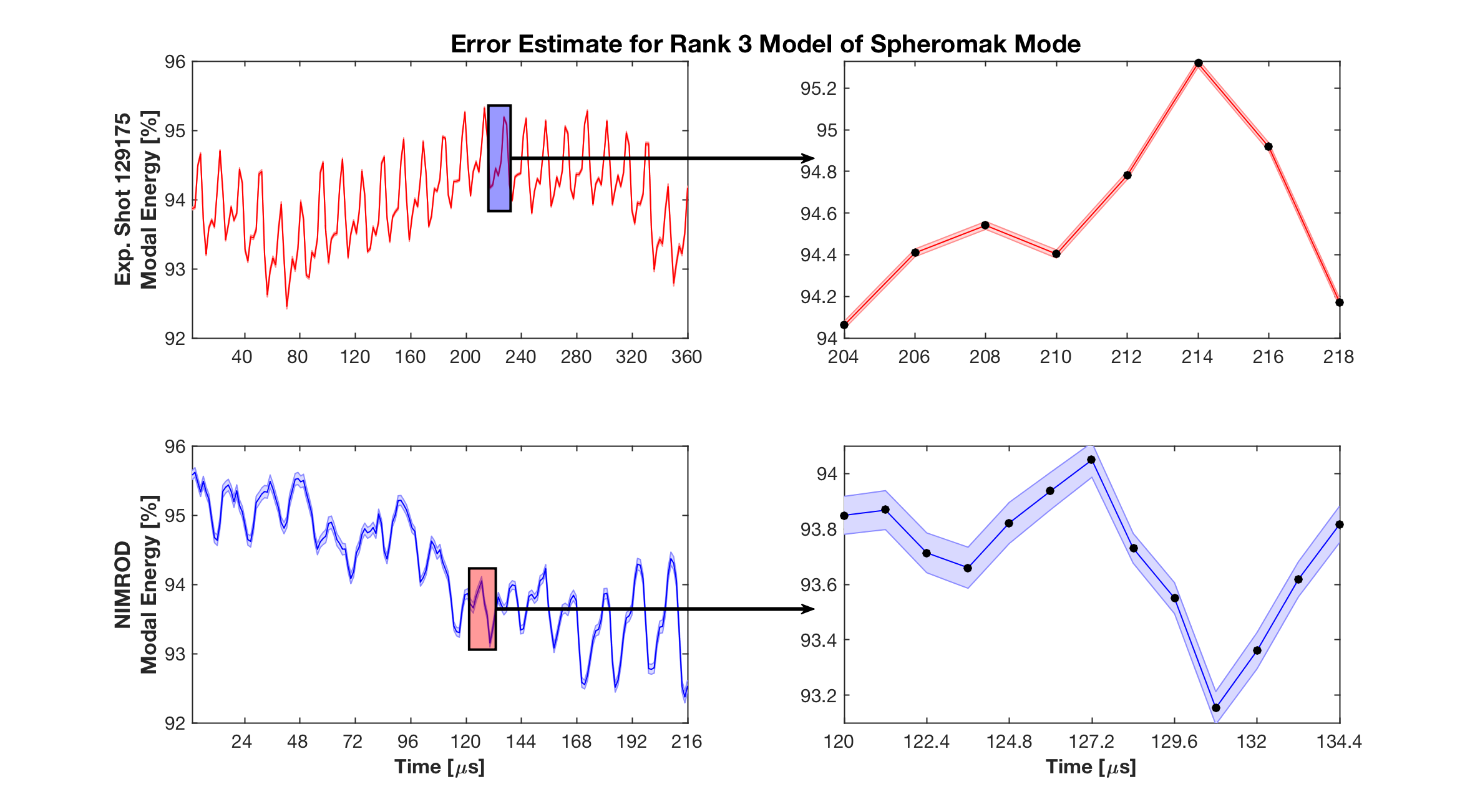} \\
\caption{Sliding-window Rank-3 DMD models exhibit very low model error for both experimental and NIMROD data sets. }
\label{fig:error_bars}
\end{figure}

In considering the accuracy of the Rank-3 model, we can quantify the model error by regarding all of the magnetic energy contained in the discarded modes as error bounds on the spheromak energy percentage. This allows an intuitive, equation-free and `physics-blind' heuristic for gauging the fit of the reduced order model. As shown in 
Fig.~\ref{fig:error_bars}, the model error on the spheromak energy is minimal.

\begin{figure}[t]
\centering
\includegraphics[scale=.25]{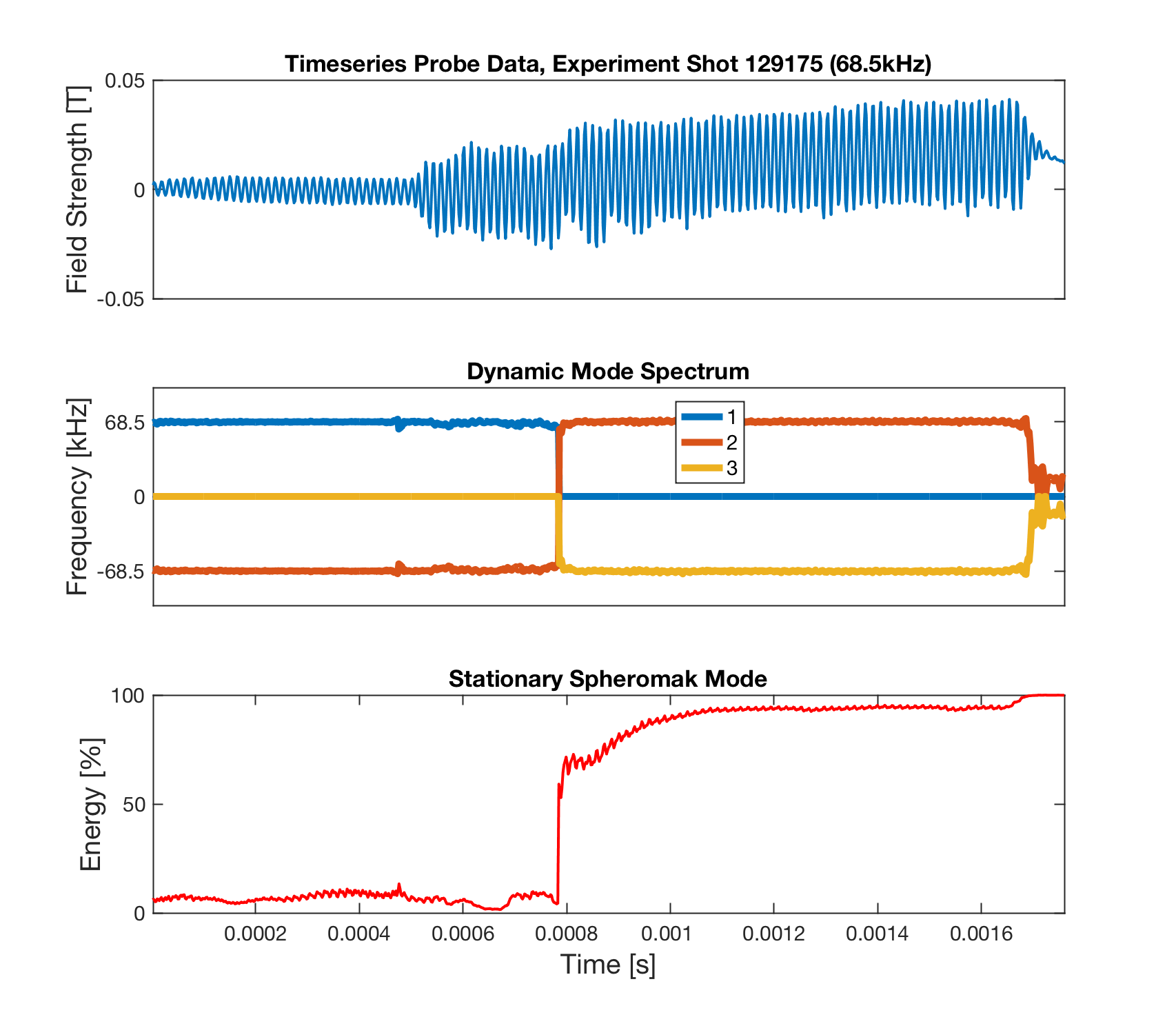} \\
\caption{With the time series data from Probe L10-225 for reference (top), the windowed DMD clearly demonstrates in both frequency (middle, \textbf{modes ranked by instantaneous energy}) and spheromak energy percent (bottom) the exact moment of spheromak formation.}
\label{fig:dmd_window}
\end{figure}

\subsection{DMD Diagnostics}

With a rank-3, 20-timestep sliding window model demonstrating very low model error, we consider the application of the DMD to experimental data for diagnostic purposes. As Fig.~\ref{fig:dmd_window} demonstrates, in a rank-3 representation the only significant temporal-spatial plasma structures are a DC magnetic field profile which resembles a Taylor-state spheromak (the \emph{spheromak mode}) and a degenerate-in-frequency mode pair representing an injector-dominated perturbation, as these two modes oscillate at or near the SIHI frequency (approximately 68.5kHz for this shot).

Running the windowed DMD across the entire experimental domain allows quick diagnosis of changes in the dynamical regime.  Just before the 400th timestep, the energy contained in the spheromak mode precipitously jumps from around 5\% to 70\%, growing to 90\% in 100 timesteps and continuing to grow until injector shut-off. This is heuristically consistent, as spheromak mode energy scales with the current gain squared (i.e.\ $E_{spheromak}/E_{inj} = I_{tor}^2 / I_{inj}^2$), and for this shot HIT-SI experienced a current gain just over 3.5. The DMD has clearly indicated the exact point in time when spheromak formation occurs by tracking the mode where $\Im[\omega] = 0$. Because the most computationally expensive component of this analysis was the singular value decomposition of a 16-by-20 matrix whose columns mostly repeat from the prior window, this could be implemented in real-time for diagnostics or control on a similar experiment.

\subsection{DMD Validation}
\begin{figure}[t]
\centering
\includegraphics[scale=.25]{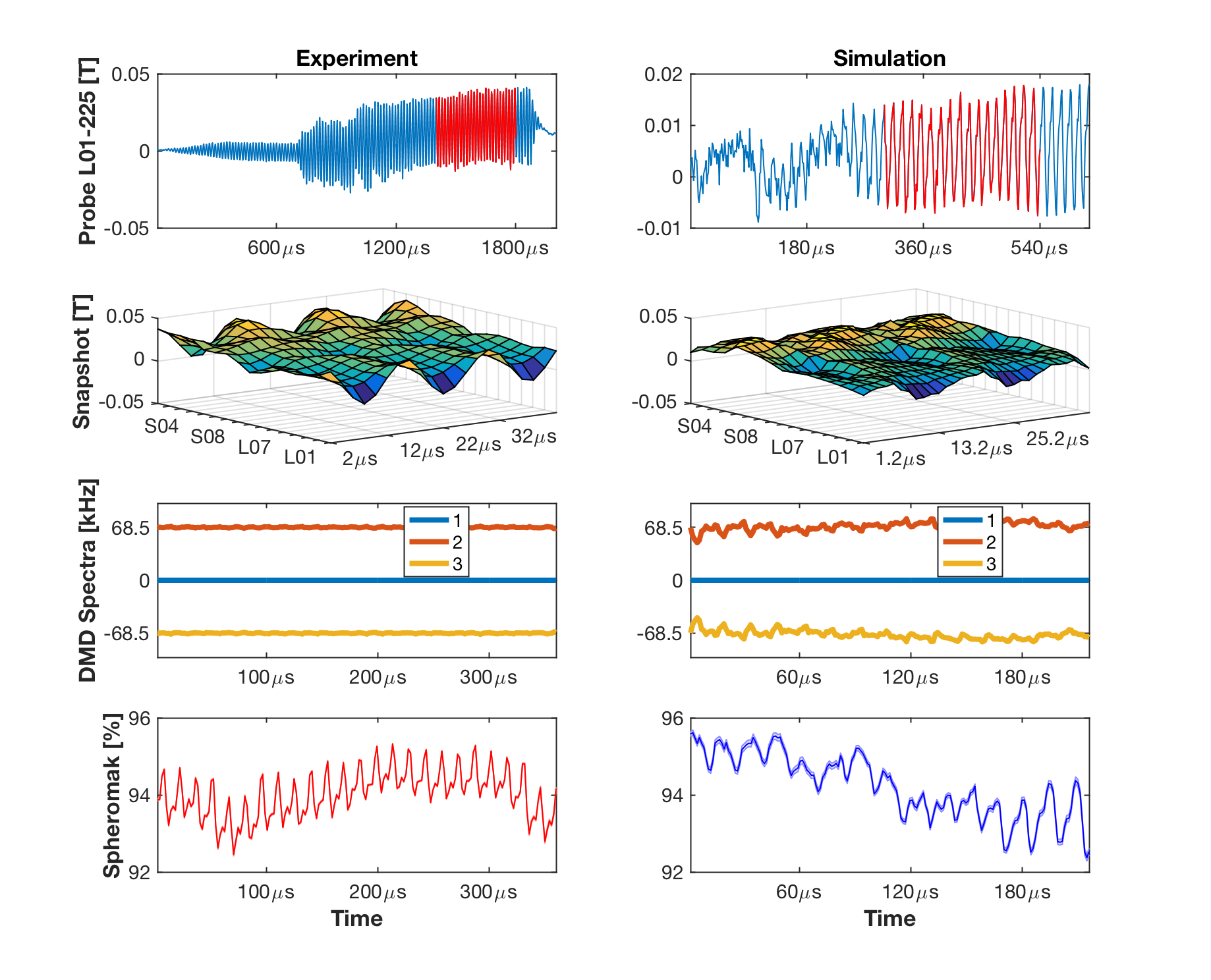} \\
\vspace*{-0.2in}
\caption{Looking at comparable dynamic regimes (top, red highlights) between NIMROD and HIT-SI, we can track and compare mode-frequency content (3rd row) and relative energy of the spheromak (bottom). A snapshot surface plot is included (2nd row) for comparison.}
\label{fig:dmd_window_freq}
\end{figure}

DMD offers a robust tool for validation of simulations against experiment. By applying the previous diagnostic approach to a comparable NIMROD output, we can directly compare the mode structure for each, as in 
Fig.~\ref{fig:dmd_window_freq}. Since our chief constraint on sampling is the Nyquist limit, which we are well above, we do not need to resample the NIMROD data onto the coarser timebase of the experiment. 

A comparison of mode frequency and energy spectra for a rank-3, windowed DMD show good agreement between NIMROD and the experiment while also highlighting key differences. In particular, we notice that in both cases the spheromak mode hovers around 94\% of total magnetic energy, while the only other dynamical modes sit nearly exactly at the SIHI frequency. Additionally, the amount of energy in the stationary spheromak mode also oscillates at the SIHI frequency. Interestingly we observed non-constant frequency behavior in the NIMROD data, which was consistent with qualitatively observed noise in this particular run. 

This technique could be generalized and expanded to any arbitrary window, provided that the matrix to be decomposed is not too highly underdetermined, to confirm a multiscale physical agreement between any arbitrary set of diagnostics and their simulated companions. We believe this to be a far more powerful tool for validation than Biorthogonal Decomposition alone, especially when paired with a principled (i.e.\ Gavish-Donoho) approach to mode truncation in a broader, multi-scale approach, eliminating the possibility of validating off of spurious or otherwise non-physical modes.

\subsection{DMD Prediction}

\begin{figure}[t]
\centering
\includegraphics[scale=.2]{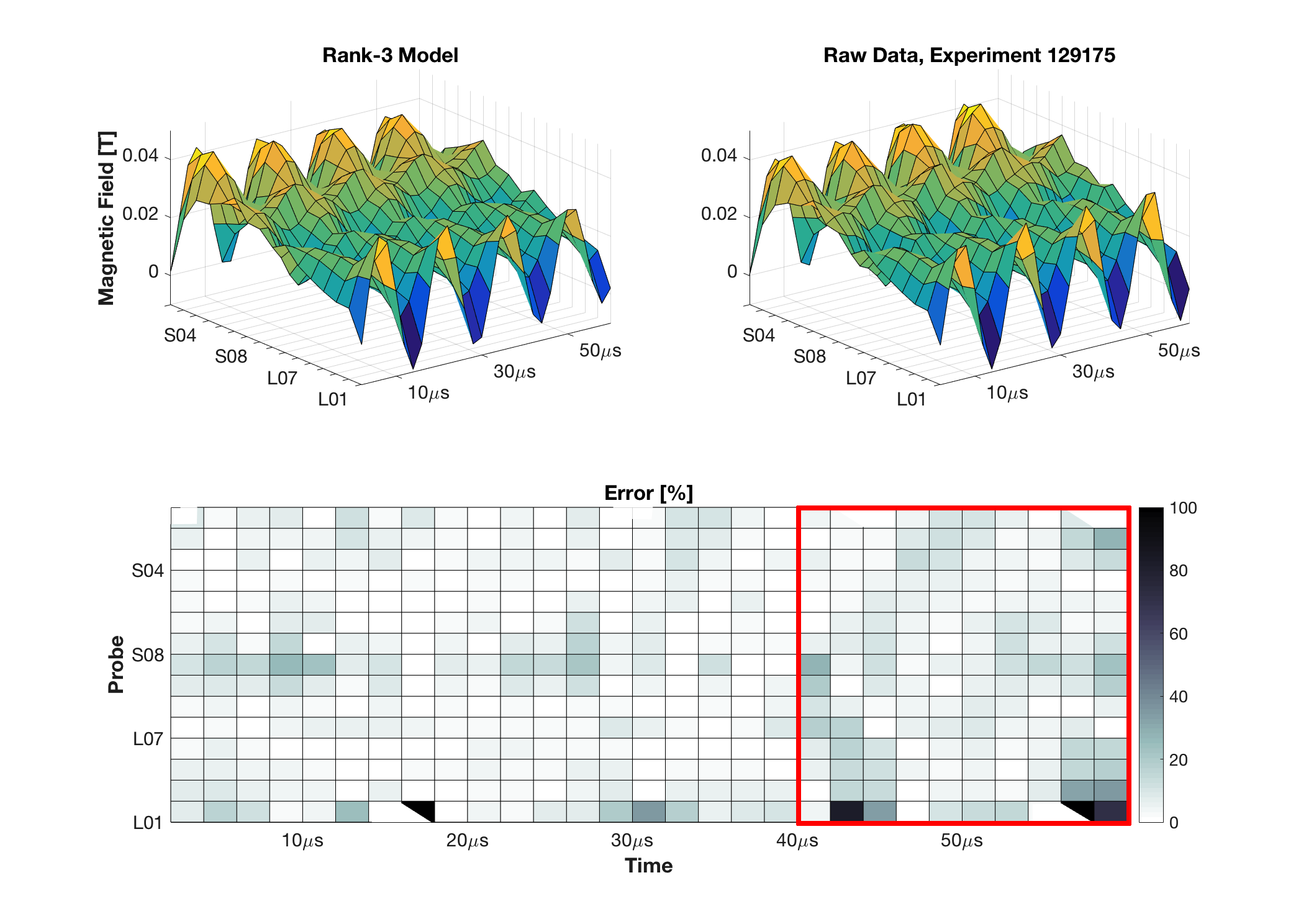} \\
\caption{Building a Rank-3 model (top-left) from \emph{the first 20 timesteps} of probe data (top-right) produces a future-state prediction nearly equivalent to truth. Error for the extrapolated region is highlighted in red (bottom).}
\label{fig:rank3_model}
\end{figure}

Finally, and as a demonstration of the technique's self-consistency, we can compute future state predictions and test the validity of the forecast. Using twenty timesteps from the experiment, we build a rank-3 dynamic mode model and allow the timebase to run ten steps into the future. These additional ten timesteps are then compared directly to the next ten timesteps from the source data. We find not only good agreement between the reduced model and the full experimental data, but also find that this agreement remains good several timesteps into the future, as shown in Fig.~\ref{fig:rank3_model}.

This computation could be run in real-time during the shot. If error began to accumulate over a user-determined threshold, the divergence from the rank-3 model would indicate the emergence of higher-order dynamics, i.e.\ of non-spurious instability modes. In this sense, future state prediction could be used as part of a dynamic mode feedback controller, though much work remains on applying this to an MHD system.  Regardless, the initial results from DMD are promising, especially in regard to thinking about DMD control methods~\cite{Proctor:2016DMDc}.

\section{Conclusions}

Reduced order models (ROMs) are of growing importance in scientific computing as they
provide a principled approach to approximating high-dimensional computations or experimental data with low-dimensional models.  Indeed, the dimensionality reduction provided by ROMs help to reduce the computational complexity and time needed to solve large-scale, engineering systems, enabling simulation based scientific studies and generating interpretable spatio-temporal features to help characterize complex systems. One of the primary challenges in producing the low-rank dynamics and spatial-temporal features is efficiently projecting the high-dimensional data to low-rank modal structures that capture both correlated spatial activity and critical time dependencies.  A variety of matrix decomposition techniques have been proposed for producing low-rank modal structures, most of them based upon the singular value decomposition.

In this work, we have proposed a new technique for obtaining reduced order models for the nonlinear dynamics of a magnetized plasma in resistive magnetohydrodynamics.  Specifically, we advocate the use of the recently developed Dynamic Mode Decomposition (DMD), an equation-free method, to decompose either computational or experimental data into spatio-temporal activity. DMD is an ideal spatio-temporal matrix decomposition that
correlates spatial features while simultaneously associating the activity with periodic temporal behavior.
With this decomposition, one can obtain a fully reduced dimensional surrogate model that can be used to reconstruct the state of the system and produce high-fidelity future state predictions. This allows for a reduction in the computational cost, and, at the same time, accurate approximations of the problem.  We demonstrate the use of the method on both numerical and experimental data, showing that it is a successful mathematical architecture for characterizing the HIT-IS magnetohydrodynamics.

The emergence of data methods like DMD across the engineering, physical and biological sciences is leading
to a host of diagnostic tools for characterizing computational and experimental complex systems.  Importantly, such
mathematical architectures also are capable of illucidating control strategies.  Indeed, the DMD method can be modified to account for inputs and output~\cite{Proctor:2016DMDc}, leading to potential data-driven controllers for the magnetohydrodynamics.  The adaptive sampling of data can also easily update the low-rank DMD models {\em on-the-fly} so as to handle parametric dependencies.   Finally, given that low-rank structures dominate the cavity dynamics, sparse sampling from limited spheromak sensors are sufficient to build the DMD dynamical models, thus leading to a purely data-driven strategy for magnetohydrodynamic characterization.  The successful application of the method on computational and experimental data attest to the efficacy of the method and its potential as an emerging data-driven diagnostic.

\section{Acknowledgements}

The authors would like to extend their gratitude to Dr. Thomas Jarboe, principal investigator of the HIT-SI experiment, for his support, wisdom, and tutelage. Furthermore, the authors would like to thank doctoral students Derek Sutherland, Chris Everson, and James Penna, undergraduate student Rian Chandra, and postdoctoral research scientist Aaron Hossack for their continual engagement and feedback. Finally, the corresponding author would like to extend his sincere gratitude for the Mary Gates Endowment at the University of Washington for financing this project.

\bibliographystyle{unsrt}
\bibliography{scibib}

\end{document}